\documentclass{article}

\usepackage[utf8]{inputenc}
\usepackage[T1]{fontenc}
\usepackage{lmodern}

\usepackage{amsmath} 
\usepackage{amssymb}
\usepackage{mathrsfs}
\usepackage{amsthm}
\usepackage{amsfonts}
\usepackage{graphicx}
\usepackage{algorithm}
\usepackage{algorithmic}
\usepackage{hyperref}
\usepackage{stmaryrd}
\usepackage{mathtools}
\usepackage{todonotes}
\hypersetup{hidelinks,breaklinks}
\usepackage{cleveref}

\usepackage{thm-restate}

\usepackage[normalem]{ulem}

\newtheorem{theorem}{Theorem}
\newtheorem{lemma}[theorem]{Lemma}

\usepackage[notion, quotation, composition]{knowledge}

\knowledge{notion}
|empty word

\knowledge{notion}
|tree
|trees
|$\Sigma$-tree

\knowledge{notion}
|arity
|arities

\knowledge{notion}
|word
|words

\knowledge{notion}
|direction
|directions

\knowledge{notion}
|alphabet
|alphabets

\knowledge{notion}
|concatenation

\knowledge{notion}
|product
|tree product

\knowledge{notion}
|branch
|branches

\knowledge{notion}
|path
|paths

\knowledge{notion}
|index

\knowledge{notion}
|parity accepting
|parity rejecting
|accepting
|rejecting
|acceptance

\knowledge{notion}
|priority
|priorities

\knowledge{notion}
|parity tree automaton
|tree automata
|automaton
|automata
|non-deterministic $I$-parity tree automaton
|$I$-automaton
|$J$-automaton
|non-deterministic automaton

\knowledge{notion}
|state
|states

\knowledge{notion}
|transition
|transitions

\knowledge{notion}
|complete

\knowledge{notion}
|transition path

\knowledge{notion}
|reachable
|reachability
|non-reachable

\knowledge{notion}
|acceptation game
|acceptation games

\knowledge{notion}
|recognized
|recognizing

\knowledge{notion}
|run
|runs

\knowledge{notion}
|accepting run
|accepting@run
|accepting runs
|rejecting run

\knowledge{notion}
|language
|languages

\knowledge{notion}
|regular tree language
|regular language

\knowledge{notion}
|$I$-feasible
|$J$-feasible
|feasible
|feasibility
|$I$-feasibility
|$J$-feasibility

\knowledge{notion}
|deterministic
|deterministic automaton
|non-deterministic

\knowledge{notion}
|guidable automaton
|guidable automata
|Guidable automata
|guidable
|guidability
|Guidability

\knowledge{notion}
|guiding function

\knowledge{notion}
|preserve acceptance
|preserves acceptance

\knowledge{notion}
|guides
|guided by
|guiding
|guided

\knowledge{notion}
|parity graph
|parity graphs
|$I$-graph
|$I$-graphs

\knowledge{notion}
|even
|odd
|even cycle
|odd cycle
|even graph
|odd graph

\knowledge{notion}
|game
|games
|parity game
|parity games

\knowledge{notion}
|play
|plays

\knowledge{notion}
|winning
|winning play
|losing
|losing play

\knowledge{notion}
|strategy
|strategies

\knowledge{notion}
|consistent with

\knowledge{notion}
|winning strategy

\knowledge{notion}
|determined
|determinacy

\knowledge{notion}
|attractor
|attractors

\knowledge{notion}
|attractor-decomposition
|attractor-decompositions
|attractor decomposition
|attractor decompositions
|decomposition
|decompositions
|($h$-height, $\kappa$-width)-attractor-decomposition

\knowledge{notion}
|tight
|tightness

\knowledge{notion}
|well-indexed

\knowledge{notion}
|tree-shape
|tree-shapes

\knowledge{notion}
|level

\knowledge{notion}
|ordered tree
|ordered trees

\knowledge{notion}
|finite tree
|finite
|finite trees

\knowledge{notion}
|child
|children

\knowledge{notion}
|node
|nodes

\knowledge{notion}
|root

\knowledge{notion}
|depth

\knowledge{notion}
|distance
|distances

\knowledge{notion}
|height
|heights

\knowledge{notion}
|diameter
|diameters

\knowledge{notion}
|subtree
|subtrees

\knowledge{notion}
|direct subtree
|direct subtrees

\knowledge{notion}
|leaf
|leaves

\knowledge{notion}
|adherence
|adherence set
|adherences

\knowledge{notion}
|height@adherence
|heights@adherence

\knowledge{notion}
|position
|positions

\knowledge{notion}
|mapping
|mappings
|mapping of $G$ into $T$
|mapping of $G$ into $T^*$

\knowledge{notion}
|offset
|offsets

\knowledge{notion}
|$n$-Strahler number
|Strahler number
|Strahler numbers

\knowledge{notion}
|universal
|universal for $\T$

\knowledge{notion}
|universal tree

\knowledge{notion}
|width

\knowledge{notion}
|isomorphically embedded
|isomorphic embedding

\knowledge{notion}
|smallest common ancestor
|smallest common ancestors

\knowledge{notion}
|$J,N$-priority transduction game
|priority transduction games
|priority transduction game
|transduction game
|transduction games
|counter
|counters
|register
|registers
|Registers

\knowledge{notion}
|transposition game
|transposition games

\knowledge{notion}
|$n$-close

\knowledge{notion}
|rank

\knowledge{notion}
|$j$-rank

\newcommand{\NN}{\mathbb{N}}

\newcommand{\A}{\mathcal{A}}

\newcommand{\C}{\mathcal{C}}

\newcommand{\F}{\mathcal{F}}
\newcommand{\G}{\mathcal{G}}

\renewcommand{\L}{\mathcal{L}}

\renewcommand{\S}{\mathcal{S}}
\newcommand{\T}{\mathcal{T}}
\newcommand{\U}{\mathcal{U}}

\renewcommand{\leq}{\leqslant}
\renewcommand{\geq}{\geqslant}

\newcommand{\ra}{\rightarrow}

\knowledgenewcommand{\concat}{\cmdkl{\cdot}}
\knowledgenewcommand{\Tr}[1]{\cmdkl{Tr_{#1}}}
\knowledgenewcommand{\trProd}{\cmdkl{\otimes}}
\knowledgenewcommand{\Lang}{\cmdkl{\L}}

\knowledgenewcommand{\attr}{\cmdkl{\mathtt{Attr}}}
\knowledgenewcommand{\restrict}{\cmdkl{\upharpoonright}}

\knowledgenewcommand{\subt}{\cmdkl{\sqsubseteq}}
\knowledgenewcommand{\h}{\cmdkl{\mathtt{h}}}
\knowledgerenewcommand{\d}{\cmdkl{\mathtt{d}}}
\knowledgenewcommand{\dist}{\cmdkl{\delta}}

\knowledgenewcommand{\Ut}[1]{\cmdkl{\U_{#1}}}

\knowledgenewcommand{\Reg}[2]{\cmdkl{\T_{#1}^{#2}}}

\knowledgenewcommand{\Succ}{\mathit{\cmdkl{Succ}}}
\knowledgenewcommand{\St}[1]{\cmdkl{\S_{#1}}}

\knowledgenewcommand{\tree}[1]{\cmdkl{T_{#1}}}

\knowledgenewcommand{\w}{\mathfrak{\cmdkl{w}}}
\knowledgenewcommand{\cw}{\mathfrak{\cmdkl{w}}}

\knowledgenewcommand{\For}{\cmdkl{\F}}

\knowledgenewcommand{\guided}{\cmdkl{\rtimes}}

\knowledgenewcommand{\AGame}[2]{\cmdkl{\G(#1,#2)}}

\knowledgenewcommand{\rk}{\cmdkl{\mathtt{rk}}}
\knowledgenewcommand{\rkj}{\cmdkl{\mathtt{rk_j}}}

\title{Revisiting the parity index hierarchy of tree automata: an informal note}
\author{Olivier Idir, Karoliina Lehtinen}

\begin{document}
	
	\maketitle

	\begingroup
	\leftskip4em
	\rightskip\leftskip
	{\bf Abstract:}
	The parity index problem of tree automata asks, given a regular tree language
	$L$, what is the least number of priorities of a nondeterministic parity tree
	automaton that recognises $L$. This is a long-standing open problem, also known
	as the Mostowski or Rabin-Mostowski index problem, of which only a few sub-cases
	and variations are known to be decidable.
	In a significant step, Colcombet and L\"oding reduced the problem to the uniform
	universality of distance-parity automata.
	In this brief note, we present a similar result, with a simplified proof, based on
	on the games in Lehtinen's quasipolynomial algorithm for parity games.
	
	We define an extended version of these games, which we call parity transduction games,
	which take as parameters a parity index $J$ and an integer bound $N$. We show
	that the language of a guidable automaton $A$ is recognised by a nondeterministic automaton
	of index $J$ if and only if there is a bound $N$ such that the parity transduction game
	with parameters $J$ and $N$ captures membership of the language, that is, for all trees $t$,
	Eve wins the parity transduction game on the acceptance parity game of $t$ in $A$ if and only
	if $t$ is in $L(A)$.
	
	\par
	\endgroup
	
\section{Introduction}

The parity index problem of tree automata asks, given a regular tree language $L$, what is the least number of priorities -- or minimal priority range $I\subset \NN$, to be exact -- such that a nondeterministic parity tree automaton with priorities in $I$ recognises $L$. This is a long-standing open problem, also known as the Mostowski or Rabin-Mostowski index problem, of which only a few sub-cases and variations are known to be decidable~\cite{SW16,Colcombet2013DecidingTW, GameAutomata,NS21,NW05}.

 However, in a significant step, in 2008, Colcombet and Löding~\cite{Guidable} reduced the index problem of a tree language to the uniform universality of a distance-parity automaton. While this reduction does not prove the decidability of the problem, it captures some deep insights into the nature of the problem and spells out the shape of a potential solution, namely finding the elusive bound.

This remarkable result, however, has a difficult, highly technical proof. In this brief note we present a similar result, with a simplified proof, which, we hope, will make the state-of-the art on this fascinating problem more accessible. 

Our proof is based on the game developed by Lehtinen for solving parity games in quasipolynomial time~\cite{RegisterGames}. 
Namely, Lehtinen reduces solving a parity game to solving a different game, in which Eve must map the original game's priorities into a smaller priority range, using a purpose-build data-structure, while guaranteeing that the sequence of outputs in this smaller range still satisfies the parity condition. Lehtinen shows that for a parity game of size $n$, the winning player can also win this harder game with output range $\mathcal{O}(\log n)$, which reduces solving the original game to solving a game of quasipolynomial size with $O(\log n)$ priorities.

Here we extend this game to the acceptance parity games of nondeterministic parity tree automata, that is, parity games with an infinite arena. We furthermore add some counters, which give  Eve some additional (but bounded) leeway in her mapping. We obtain a game that we call the parity transduction game $\Reg{J}{N}(G)$, played over a parity game $G$, with parameters $J$, the output priority range, and $N$, the bound on the counters.

We show that the $J$-feasibility of the language of a guidable automaton $\A$ is characterised by the existence of an interger $N$ such that the parity transduction game with parameters $J$ and $N$ coincides with the acceptance game of $\A$, written $\AGame{\A}{t}$ for all input trees $t$.

\begin{restatable}{theorem}{thmfeasibilitygame}\label{thm:feasible-register}
Given a "guidable automaton" $\A$, $J$ an "index", the following are equivalent:
\begin{itemize}
\item $\Lang(\A)$ is $J$-"feasible".
\item There exists $n\in \NN$ such that for all $\Sigma$-"tree" $t$, $t\in \Lang(A)$ if and only if Eve wins $\Reg{J}{N}(\AGame{\A}{t})$
\end{itemize}

\end{restatable}
 
For the forward implication, we use the guidability of $\A$: a run of the hypothetical $J$-automaton recognising $L$ guides both Eve's run and her mapping of priorities into the range $J$. A simple pumping argument shows the correctness of this strategy.

For the backward implication, we construct an automaton that encodes the transduction game using priorities in the range $J$ by encoding the registers and counters (bounded by the appropriate $N$) in the state-space. Then composing this automaton with $\A$ gives the equivalent $J$-automaton.

Towards the correspondence between our result and the Colcombet-L\"oding reduction, one can obtain a distance-parity automaton by encoding the values of the counters into the distance function.

\section{Preliminaries}\label{sec:definitions}

The set of natural numbers $\{0, 1, \dots\}$ is denoted $\NN$, the set of strictly positive numbers is denoted $\NN^+$. An ""alphabet"" is a finite non-empty set $\Sigma$ of elements, called letters (or ""directions"" when the alphabet is $\{0,1\}$).
$\Sigma^*$ and $\Sigma^\omega$ denote the sets ot finite and infinite ""words"" over $\Sigma$, respectively.
For $u$ a (possibly infinite) "word" and $n\in \NN$, the "word" $u|_{n}$ consists of the first $n$ symbols of $u$. For $u$ and $v$ finite "words", $u \concat v$ denotes the ""concatenation"" of $u$ and $v$.
The ""empty word"" is denoted $\varepsilon$. The length of a finite "word" $u$ is written $|u|$.

\AP An ""index"" $[i,j]$ is a non-empty finite range of natural numbers $I = \{i, i+1,\dots, j\} \subseteq \NN$.
Elements $c \in I$ are called ""priorities"". We say that an infinite sequence of "priorities" $(c_n)_{n\in \NN}$ is ""parity accepting"" (or simply \reintro*"accepting") if $\limsup_{n \to \infty} \equiv 0 \mod 2$, else it is "parity rejecting" (or "rejecting").

\subsection{Parity games}
\AP For $I$ an "index", $(V,E)$ a graph with $V$ a countable set of vertices and $L:E\to I$ an edge labeling, we call $G = (V,E,L)$ a ""$I$-graph"", or a "parity graph". We say that a (possibly infinite) path through $G$ is ""even"" if the maximal priority on its edges is even. Else, it is ""odd"". The graph is said \reintro{even} if all its infinite paths are "even".

\AP A ""parity game"" played by players Eve and Adam consists in a "parity graph" $\G = (V,E,L)$ with a partition of $V$ in two sets: $V = V_E \sqcup V_A$, controlled respectively by Eve and Adam, and with $E$ such that all vertices have an outgoing edge. A ""play"" of $\G$ starting in $v\in V$ consists in an infinite sequence of edges $\rho := (e_i)_{i\in \NN}$ forming an infinite path starting in $v$. 
A "play" $(e_i)_{i\in \NN}$ is ""winning"" for Eve (or simply \reintro{winning}) if $(L(e_i))_{i\in \NN}$ is "parity accepting", else it is said to be ""losing"" (for Eve, and "winning" for Adam).

\AP A ""strategy"" for Eve consists of a function $\sigma : E^* \to E$ such that, for all "play" $\rho$, for all $n \in \NN$, if $\rho_{|n}$ ends in a vertex $v \in V_E$, $\sigma(\rho_{|n})$ follows an edge from $v$. A "play" $\rho$ is said to be ""consistent with"" the strategy $\sigma$ if for all $n$, $\rho_{|n}$ ending in a vertex of $V_E$ implies that $\rho_{|n+1} = \rho_{|n}\sigma(\rho_{|n})$. 
We say that a Eve "strategy" $\sigma$ is ""winning@winning strategy"" from vertex $v\in V$ if all plays "consistent with" $\sigma$ starting in $v$ are "winning". We similarly define "strategies" for Adam, winning when all plays "consistent with" them are "winning" for Adam.

\AP A notable result on parity games is their ""determinacy"": either Eve has a winning strategy or Adam does, and the two situations are exclusive\cite{BorelDeterminacy}.

A "strategy" for Eve in a game $\G = (V_E\sqcup V_A,E,L)$ induces an Adam-only "game" $\G'$ played on the unfolding of $\G$, from which are removed all the edges that Eve does not choose. This game can be seen as a "parity graph", as the partition of the vertex set is now a trivial one, which is "even" if the "strategy" is "winning@winning strategy".

\subsection{$\Sigma$-trees and automata}
\AP A ""$\Sigma$-tree"" (or directly \reintro*{tree}) is a function $t : \{0,1\}^* \to \Sigma$. The set of all $\Sigma$-labelled trees is denoted ${\Tr \Sigma}$.

\AP An infinite sequence of directions $b \in \{0,1\}^\omega$ is called a ""branch"".
Given a  "tree" $t\in \Tr\Sigma$, a ""path"" $p$ (along a "branch" $b$) is a sequence $(p_i)_{i \in \NN} := (t(b_{|i}))_{i\in \NN}$.

\AP A ""non-deterministic $I$-parity tree automaton"" (also called \reintro*"$I$-automaton", or "automaton" of "index" $I$) is
a tuple $A = (\Sigma, Q_A , q_{i,A}, \Delta_A, \Omega_A )$, where $\Sigma$ is an alphabet, $Q_A$ a finite set of ""states"", $q_{i,A} \in Q_A$ an initial "state", $\delta \subseteq Q_A \times \Sigma \times Q_A \times Q_A$ a transition relation; and $\Omega_A : \Delta_A \to I^2$ a "priority" mapping over the edges. 
A ""transition"" $(q, a, q_0 , q_1) \in \Delta_A$, is said to be from the "state" $q$ and over the letter $a$. By default, all "automata" in consideration are ""complete"", that is, for each state $q \in Q_A$ and letter $a\in \Sigma$, there is at least one "transition" from $q$ over $a$ in $\Delta_A$.
When an automaton A is known from the context, we skip the subscript and write just $Q,\Delta$, etc.

\AP For $q,q' \in Q$, a ""transition path"" from $q$ to $q'$ corresponds to a finite "transition" sequence $(q_j, a_j, q_{j,0}, q_{j,1})_{j< N} \in \Delta^N$ such that $q = q_0$, and $\forall j < N, q_{j+1}\in \{q_{j,0}, q_{j,1}\}$ with $q_{j+1} = q'$.

\AP Given a tree $t \in \Tr\Sigma$, and an $I$-"automaton" $A$, the ""acceptation game"" of $A$ on $t$, also denoted $\intro*\AGame{A}{t}$, is the "parity game" obtained by taking the product of $A$ and $t$. More precisely, its arena consists in $\{0,1\}^*\times (Q_A \cup \Delta_A)$, where all the states of the shape $\{0,1\}^*\times Q_A$ are controlled by Eve, and the others by Adam. 
\begin{itemize}
	\item When in a state $(w,q) \in \{0,1\}^* \times Q_A$, Eve chooses a transition $e\in \Delta_A$ of the shape $(q,t(w),q_0,q_1)$, and the game moves to the state $(w,e)$. All these transitions have for label the minimal priority in $I$.
	\item Let $q\in Q_A$ and $e=(q,a,q_0,q_1) \in \Delta_A$. In a state $(w,e)$, Adam chooses either $0$ or $1$, and the games then moves towards either $(w\concat 0, q_0)$ or $(w\concat 1, q_1)$. For $\Omega_A(e) = (i_0,i_1)$, these transitions have respectively for priority $i_0$ and $i_1$.
\end{itemize}
\AP We say that $t$ is ""recognized"" by $A$ if Eve wins $\AGame{A}{t}$. The set of "trees" "recognized" by $A$ is called the ""language"" of $A$ and is denoted $\intro*\Lang(A)$.

\AP If we fix a "strategy" for Eve, the "acceptation game" becomes an Adam-only game, called a ""run"" of $A$ on $t$. We observe that it is played on a "parity graph" of the shape of a binary tree. We thus observe that a "run" can be considered as a tree in $\tree{\Delta_A}$. This "run" is won by Adam if and only if there exists a "parity rejecting" "branch". In this case, it called a ""rejecting run"", else it is an ""accepting run"".\\
Up to forgetting the exact edges and neglecting the first priority (which we can do as the parity condition is prefix-independant), we can map this "run" towards a tree in $\tree{I}$, and it remains "accepting" on the same condition.

\AP A set of trees $L \subseteq \tree{\Sigma}$ is a ""regular tree language"" if it is of the form $\Lang(A)$ for some "automaton" $A$. It is said "$I$-feasible" if furthermore $A$ is of "index" $I$.

\subsection{Guidable automata}

The notion of a "guidable automata" was first introduced in \cite{Guidable}.
They can be conceived as a form of history-determinism applied to "tree automata". These "automata" are not necessarily determinisable: for instance, the language of "trees" that have an \texttt{a} somewhere is a "regular language", recognized by a simple two-state nondeterministic reachability automaton. Yet there does not exist a deterministic automaton recognizing this language, as it would need to know in advance in which direction to look for the \texttt{a}.\\
"Guidability" circumvents this issue: instead of guessing directions, a "guidable automaton" takes as input an accepting run of another "automaton" recognizing a sublanguage, and uses it to solve its nondeterminism. Intuitively, they are "automata" that fairly simulates all "automata" recognizing this language. "Guidable automata" are fully expressive \cite[Theorem 1]{Guidable} and are more manageable than fully nondeterministic automata. 

\AP Fix two "automata" $A$ and $B$ over the same "alphabet" $\Sigma$. A ""guiding function""
from $B$ to $A$ is a function $g : Q_A \times \Delta_B \to \Delta_A$ such that $g(p, (q, a, q_0 , q_1 )) = (p, a, p_0 , p_1)$
for some $p_0, p_1 \in Q_A$ (i.e. the function g is compatible with the "state" $p$ and the letter $a$). 

If $\rho \in Tr_{\Delta_B}$ is a "run" of $B$ over a "tree" $t \in \Tr\Sigma$ then we define the "run" $g (\rho) \in Tr_{\Delta_A}$ as follows. We define inductively $q : \{0,1\} \to Q_A$ in the following fashion: $q(\varepsilon) = q_{i,A}$, and supposing $q(u)$ to be defined, for $\rho(q(w),\rho(w))=(q(w),t(w),q_0,q_1)$, we let $q(u\concat 0),q(u\concat 1)$ to be respectively $q_0,q_1$. We can then define the run $g (\rho) \in \tree{\Delta_A}$ as 
$$g(\rho): u \mapsto g(q(u),\rho(u)).$$
Notice that directly by the definition, the "tree" $g(\rho)$ is a "run" of $A$ over $t$. 

\AP We say that a "guiding function" $g : Q_A \times \Delta_B \to \Delta_A$ ""preserves acceptance"" if whenever $\rho$ is an "accepting run" of $B$ then $g(\rho)$ is an "accepting run" of $A$. We say that an "automaton" $B$ ""guides"" an "automaton" $A$
(denoted $B \ra A$), if there exists a "guiding function" $g : Q_A \times \Delta_B \to \Delta_A$ which "preserves acceptance". In particular, it implies that $\Lang(B) \subseteq \Lang(A)$.

\AP An "automaton" $A$ is ""guidable"" if it can be "guided by" any "automaton" $B$ such that $L(B) = L(A)$
(in fact one can equivalently require that $L(B) \subseteq L(A)$, see \cite[, remark 4.5]{lodingHDR}).
We will use the following fundamental theorem, stating that "guidable automata" are as expressive as "non-deterministic" ones.

\begin{theorem}\cite[, Theorem 1]{Guidable}
	For every regular tree language $L$, there exists a "guidable automaton" "recognizing" $L$. Moreover, such an "automaton" can be effectively constructed from any "non-deterministic automaton" "recognizing" $L$.
\end{theorem}

The following is a useful lemma stating that given a pair of "runs" where one is "guided by" the other, the preservation of global acceptance implies also a more local version: between all pumpable pairs of states, that is, pairs of states that are not distinguished by either run, if the "guiding" run is dominated by en even priority, then so is the "guided" one.
\begin{lemma}\label{cl:pump}
	Let $A,B$ be "automata", let $t\in \Tr{\Sigma}$, let $\rho_A, \rho_B$ be "accepting runs" over $t$ of $A$ and $B$, respectively, where $A$ is "guided by" $B$. We consider these "runs" as trees in $\Tr{\Delta_A},\Tr{\Delta_B}$ respectively. 
	Given $u,v \in \{0,1\}^*, \{0,1\}^+$ such that $\rho_A(u) = \rho_A(u\concat v)$, and such that $\rho_B(u) = \rho_B(u\concat v)$, if the greatest priority encountered between positions $u$ and $u \concat v$ in $\rho_B$ is even, so is the greatest priority encountered in this segment in $\rho_A$.
\end{lemma}
\begin{proof}
	Given an infinite "tree" $t$, we define the "tree" $t^*$ starting from $t$, where, for $t_u$ the subtree of $t$ at position $u$, for all $n \in \NN$, $t_u$ replaces recursively the subtrees at positions $u \concat v^n$.\\
	We use the same construction to define the runs $\rho_A^*$ and $\rho_B^*$. They are legitimate "runs" on $t^*$, as $\rho_B(u) = \rho_B(u\concat v)$, therefore at each repetition of $t_u$, $A$ is in the same state $q_u$ and can thus choose the same transition. The same applies to $\rho_B$.
	
	Let $b^*$ the unique "branch" going through all the repetitions of $t_u$ in $\rho_B^*$. We observe  that this corresponds to the "branch" $u \concat v^\omega$. Therefore, past the position $u$, $b^*$ infinitely repeats the segment from $u$ to $u\concat v$ of the subtree $\rho_B(u)$. This segment is dominated by an even priority $p$ by lemma hypothesis. Therefore $b^*$ is dominated by $p$ even, and is thus "accepting@accepting run".
	
	We observe that on  any other "branch" $b$ of $\rho_B^*$ (that is, $b$ is of the shape $u\concat v^k \concat w$ with $w \neq v^\omega$), the suffix of "path" along $b$ in $\rho_B^*$ is a clone of the "path" along $u\concat w$ in $\rho_B$, and is thus "accepting". We thus obtain that $\rho_B^*$ is accepting.
	
	Let us denote $g$ the "guiding function" $g : Q_A \times \Delta_B \to \Delta_A$, $(q_i)_{i < |v|}$ the states taken by $A$ between $u$ and $u \concat v$, and $(\delta_i)_{i < |v|}$ the transitions taken in $B$ between $u$ and $u \concat v$. We have, as $\rho_A$ is guided by $\rho_B$, that on this (repeated) segment, $\rho_A$ takes the transitions $(g(q_i, \delta_i))_{i< |v|}$.
	Hence, as $q_0$ is repeated at position $u\concat v$, and in $\rho_B'$ the same transitions are repeated along this "branch", we obtain that the "run" $\rho_A^*$ is a "run" guided by $\rho_B^*$. Notably, as $\rho_B^*$ is "accepting@accepting run", so is $\rho_A^*$: therefore, on the "branch" $b^*$ along $u \concat v^\omega$, $\rho_A^*$ is dominated by an even priority, hence an even priority dominates $(q_i)_{i<|v|}$.
\end{proof}

\section{Game characterisation of the parity index}\label{sec:register-game}

In this section we define "priority transduction games", based on the register games from Lehtinen's algorithm  in~\cite{RegisterGames}, augmented with some counters. We characterise the "$J$-feasibility" of a "language" $L(A)$, where $A$ is a "guidable automaton", by the existence of a uniform bound $N\in \mathbb{N}$ such that a "tree" is in $L(A)$ if and only Eve wins the $J$-"priority transduction game" on $\AGame{A}{t}$, with counters bounded by $N$.

The idea of these "priority transduction games" is that in addition to playing the "acceptation game" of an $I$-automaton $\A$ over a tree $t$, which has priorities in $I$, Eve must map these priorities on-the-fly into the "index" $J$. In the original games from~\cite{RegisterGames}, she does so by choosing at each turn a register among roughly $\frac{J}{2}$ registers. Each register stores the highest priority seen between two turns where Eve choses it. Then, the output is a priority in $J$ which depends on both the register chosen and the parity of the value stored in it. 
Here, the mechanism is similar, except that we additionally have counters that allow Eve to delay outputting odd priorities a bounded number of times.

\AP Formally, for $J$ a priority "index" (of minimal value assumed to be $1$ or $2$ for convenience), $N \in \NN$, the ""$J,N$-priority transduction game"" is a game outputting priorities in $J$, played by Eve and Adam, over an $I$-"parity graph" $G=(V,E,L)$ for $I$ an "index". Such a game has two parameters, $J$ the output "index", and $N$ the bound of its "counters", and is denoted $\intro{\Reg{J}{N}}(\rho)$. A configuration of the game corresponds to a position $p\in V$, a value in $I$ for its \reintro*"registers" $r_j$ for all even $2j\in J$ (if $1 \in J$, it has an additionnal "register" $r_0$), and a value between $0$ and $N-1$ for all its \reintro*"counters" $c_{i,j}$ with $i$ odd $\in I, j$ such that $r_j$ is a "register".\\
Starting from some initial vertex $p_0\in V$ with "counters" set to $0$ and "registers" set to $min(I)$, the game proceeds as follows at step $l$ :
\begin{itemize}
	\item Adam chooses an exiting edge $e = (p,p')\in E$ ; the position becomes $p'$.
	\item Eve chooses a "register" $r_j$
	\item The game produces the output $w_l$ :
		\begin{itemize}
		\item if $j = 0$, $w=1$ (recall that $r_0$ is a "register" iff $1 \in J$). Else,
		\item if $r_j$ is even, $w=2j$
		\item if $c_{r_j,j} = N$, it is said to reach $N+1$ before being reset: $w_l=2j+1$ and $c_{r_j,j} := 0$. If $2j+1 \notin J$, Eve loses instantly.
		\item else, $w_l = 2j$ and $c_{r_j,j} := c_{r_j,j} +1$
		\end{itemize}
	\item If $L(e)$ is even, we pose $i:= L(e)$, else Eve chooses an odd $i$ such that $L(e)\leq i$ (choice $\sharp$\label{sharp}). 
	Then the following updates occur : 
	\begin{itemize}
		\item Smaller "counters" are reset : $\forall i' < i, c_{i',j} := 0$ and $ \forall j'<j, c_{r_j,j'} := 0$,
		\item "Registers" get updated : $\forall j' > j, r_{j'} := \max(i, r_{j'})$, and $r_j := i$
	\end{itemize} 
\end{itemize}
Eve wins if the infinite sequence of outputs $(w_l)_{n\in \NN}$ is "parity accepting", else Adam wins.\\

Intuitively, the "counters" correspond to a kind of error margin for Eve. She has for input a sequence of $i\in I$, which she wants to convert into an accepting sequence of $(w_l)_{l\in \NN}\in J$. She only outputs an odd $w_l$ after some "counter" $c_{i,j}$ reaches $N+1$, that is, if she chose $N+1$ times the "register" $r_j$ while input sequence dominated, between each such choice, by some odd $i$. Indeed, in this case, $r_j$ has for value some $i'$ just after being chosen, but as $r_j$ is non-decreasing until is is chosen again, it will eventually take the value $i\geq i'$, as it dominates the input sequence between each $r_j$ choice.

The fact that "registers" are non-decreasing implies that if the input is dominated by some odd $i$, Eve cannot produce an accepting sequence by merely choosing a small $j'$ when witnessing $i$: all the "registers" $r_{j}$ for $j'\leq j$ will take for value $i$, and thus any greater "register" chosen afterwards will remember seeing that $i$.

The "register" $r_0$ allows Eve to possibly wait some time, for instance if the current input sequence would make her loose instantly, though she looses if she waits indefinitely.
Finally, the choice $\sharp$ allows her to break a sequence dominated by many identical $i$ in a sequence with some greater $i'$ in between, resetting the "counters" albeit at the cost of witnessing a greater odd priority.

Let $G$ be a "parity game" of "index" $I$, $N\in \NN$, $J$ an "index". We define the game $\Reg{J}{N}(\G)$ as the game $\Reg{J}{N}$ where, instead of following an Adam-only game $G$, it follows an ongoing play of $G$ where the player owning the current position $q$ chooses its move in $\G$ at each step, before Eve chooses her "register". It corresponds to the composition of $\Reg{J}{N}$ with the game $\G$. If we fix a strategy $\sigma$ for Eve in $\G$, we observe $\Reg{J}{N}(\G)$ corresponds exactly to the "priority transduction game" $\Reg{J}{N}$ over the Adam-only game induced by $\sigma$, and that Eve wins $\Reg{J}{N}(\G)$ if and only if she wins $\Reg{J}{N}(\G_\sigma)$, for $\G_\sigma$ the "parity graph" induced by some "strategy" $\sigma$ in $\G$.

Note that whatever its input, $\Reg{J}{N}$ corresponds to a parity game, and is therefore "determined".

We now show that these games are intrinsically linked to the "index" required to recognize a "regular tree language".

\thmfeasibilitygame*


In order to prove this theorem, we first show that Eve can only win $\Reg{J}{N}(G)$ for $G$ an "even@even graph" "parity graph"
\begin{lemma}\label{lem:reg-rejects-rejecting}
	Let $G$ a "parity graph". If $G$ is not "even@even graph", then for all $J,N$, Adam wins $\Reg{J}{N}(G)$
\end{lemma}
\begin{proof}
	As $G$ is not "even@even graph", there exists an infinite "path" in $G$ dominated by some odd $\hat{i}$.
	In $\Reg{J}{N}(G)$, Adam will simply follow this path. Therefore, the sequence $L(e)$ is dominated by $\hat{i}$, and thus the sequence of $i\in I$ chosen by Eve is dominated by some odd $i^*\geq \hat{i}$. This $i^*$ is therefore maximal among all "priorities" seen after the step $n_0$, for some $n_0 \in \NN$.\\
	Let $(j_n)_{n\in \NN}$ the infinite sequence of "registers" chosen by Eve, dominated by some $j^*$, therefore maximal among all "register" indices chosen after some $n_1 \geq n_0 \in \NN$. Let us look at events past the $n_1$-th step of the game.\\
	If $j^* = 0$, the game only outputs $\min(I)$, odd, and Adam indeed wins. We can thus restrict ourselves to the case were $j^* \neq 0$.
	Infinitely often, $\rho(p) = i^*$, hence, for $j_0$ picked at the corresponding step, $\forall j' \geq j_0, r_{j'}\geq i^*$. It is notably the case for $r_{j^*}$, by maximality of $j^*$. Therefore, infinitely often, as a $j^*$ will recur and that $r_{j^*}$ cannot be reduced before then (as the value of a "registers" is non-decreasing until it is chosen), we will see $r_{j^*} = i^*$. In which case, we end in one of the two latter case of the case disjunction : either we output $2j^*+1$, either we increment $c_{i^{*},j^*}$. Along the "branch" $b$ we no longuer see a value superior to $i^*$, nor pick a "register" superior or to $j^*$. Hence, the "counter" $c_{i^*,j^*}$ is never reset : we thus output infinitely often a $2j^{*}+1$ (or even immediately lose if $2j^* = \max(J)$).
	As we no longer pick any "register" $> j^*$, we easily see that we never output any "priority" $>2j^*+1$, which concludes as to the fact that the output sequence is "rejecting".
\end{proof}

The following lemma is key to proving the correctness of the construction. The idea behind it is Eve using the guidability by a run $\rho_B$ on an equivalent $J$-"automaton" to choose her "registers", and we establish that it indeed defines a winning strategy on $\Reg{J}{N}(\rho_A)$ for $\rho_A$ accepting run of $A$ guided by $\rho_B$.

\begin{lemma}
	Let $A$ a guidable automaton. If $\Lang(A)$ is $J$-feasible witnessed by an "automaton" $B$, for $N = |A| |B|+1$, for all $\Sigma$-tree $t \in \Lang(A)$, for $\rho_A$ the "run" of $A$ on $t$ "guided by" an "accepting run" $\rho_B$ of $B$ over $t$, Eve wins $\Reg{J}{N}(\rho_A)$.
\end{lemma}
\begin{proof}
	We consider the "runs" as "trees" in $\Tr{\Delta_A}$ and $\Tr{\Delta_B}$ respectively. That is, the choice made by Adam at each step amounts to choosing a direction in $\{0,1\}$.\\
	We first describe the strategy for Eve in $\Reg{J}{N}(\rho_A)$. We once again suppose for convenience that $\min(J)\in \{1,2\}$. At position $p$, where $\rho_B(p)\in [2j, 2j+1]$ (for some $j$ such that $2j\in J$), Eve chooses the "register" $r_j$. If $\rho_B(p) = \min(J) $, Eve chooses $r_0$. She always chooses $i = \rho(p')$ for the choice $\sharp$. This strategy being fixed, let us verify that Eve wins whatever Adam plays, that is, for all "branches" $b$.\\
	As $\rho_B$ is "accepting", its maximal infinitely recurring priority along $b$ is even, let's say $2j^{*} \in J$ (therefore $j^* \neq 0$, as $0 \notin J$). Due to "guidability", $\rho_A$ is also dominated by some even $i^*$ along $b$. We look past the point where we no longer see any priority superior to either, in their respective "trees".
	Then, as in the above proof, infinitely often, $r_{j^*}=i^*$ when the "register" $j^*$ is picked. Which means that at these (infinitely recurring) moments, $\Reg{k}{N}$ outputs a $2j^*$. Let us show that it never outputs any greater "priority" (nor instantly loses).\\
	As we see no priority in $\rho_B$ greater than $2j^{*}$, the "register" of greatest index picked is $r_{j^*}$, hence the greatest "priority" that could be output is $2j^*+1$ (or, if $2j^* = \max(J)$, Eve instead instantly looses). If by contradiction either happened, then for some $i\in I, c_{i,2j^*}$ reaches $N$ without being reset. That is, we would have along $b$ a contiguous subsequence in $(\rho_A, \rho_B)$ passing at least $N+1$ times by an edge of "priority" $i$ odd in $\rho_A$, then (or possibly at the same time) an edge of priority $2j^*$ in $\rho_B$, $i$ and $2j^*$ respectively dominating this whole subsequence in $\rho_A$ and $\rho_B$. As $N> |A \times B|$, along our progression through $b$, there is a segment starting in a position $u$ and ending in a position $u\concat v$ such that $\rho_A(u) = \rho_B(u) = \rho_A(u\concat v) = \rho_B(u \concat v)$, comprising at least one occurence of $i$ and at least one occurence of $2j^*$, both dominating in their segment of respectively $\rho_A$ and $\rho_B$. This is impossible as, according to lemma \ref{cl:pump}, such an $i$ then needs to be even. Therefore $c_{i, 2j^*}$ never reaches $N+1$, and $2j^*$ indeed dominates the output of $\Reg{k}{N}$ (or, if $2j^* = \max(J)$, Eve does not loose instantly) : the sequence is therefore "accepting".
	
\end{proof}

By combining the two previous lemmas, we prove the direct implication of Theorem \ref{thm:feasible-register}. Finally, we prove the converse implication, composing $A$ with a "$J$-automaton" recognizing trees on which Eve wins $\Reg{J}{N}$.
\begin{lemma}
	Let $J$ a priority "index", let $A$ be a "guidable automaton" such that there exists $n\in \NN$ such that for all $\Sigma$-tree $t$, $t\in \Lang(A)$ if and only if Eve wins $\Reg{J}{N}(\AGame{A}{t})$, then there exists an "automaton" of "index" $J$ such that $\Lang(B)=\Lang(A)$.
\end{lemma}
\begin{proof}
	Let $C$ an automaton describing the "transduction game" $\Reg{J}{N}$, that is, accepting "trees" $\rho$ such that Eve wins $\Reg{J}{N}(\rho)$. 
	That is, $C$ has for states the set $Q_C$ the set of the different configurations of $\Reg{J}{N}$, which are in finite number, and on an input tree $\rho$, $C$ behaves like $\Reg{J}{N}(\rho)$. Formally, for $i\in I$ (the label of the current position in $\rho$) and a configuration $c$, let $\C$ the sets configurations that can be reached by Eve in one step in $\Reg{J}{N}$ (from the point where Adam made his move, bringing us to the position of label $i$). Then $\delta_C(c,i) = \C$, each transition being labelled with the corresponding output $w \in J$. We observe that on each branch $b \in \{0,1\}^\omega$, the different "runs" in $C$ correspond to the different plays of $\Reg{J}{N}(\rho)$ where Adam successively chose the directions of $b$.
	Let us design $B$ as the composition of $A$ and $C$. That is, $B$ takes for input some $\Sigma$-"tree" $t$, on which $A$ admits a "run" $\rho$. Then $C$, taking $\rho$ as input, accepts if Eve wins $\Reg{J}{N}(\rho)$. We observe that $B$ has states $(q,c) \in Q_A \times Q_C$, hence, as $Q_C$ is finite, this composition indeed forms an "automaton" of "index" $J$. Let us show that it recognizes exactly $\Lang(A)$.\\
	If $t\notin \Lang(A)$, all the "runs" of $A$ on $t$ are "rejecting", hence by lemma \ref{lem:reg-rejects-rejecting}, the output of $\Reg{J}{N}$ is rejecting on any "run" of $A$ on $t$. Else, $t\in \Lang(A)$, and by hypothesis, Eve wins $\Reg{J}{N}(\AGame{A}{t})$. Therefore, for $\rho_t$ the "run" of $A$ that she uses to win $\AGame{A}{t}$, Eve can follow her strategy in $\Reg{J}{N}(\AGame{A}{t})$ to resolve the non-determinism of $C$. Therefore $t \in \Lang(B)$. Hence, $\Lang(B) = \Lang(A)$, and $\Lang(A)$ is $J$-"feasible".
\end{proof}

\bibliographystyle{plain}
\bibliography{biblio}

\begin{thebibliography}{1}

\bibitem{Colcombet2013DecidingTW}
Thomas Colcombet, Denis Kuperberg, Christof L{\"o}ding, and Michael~Vanden
  Boom.
\newblock Deciding the weak definability of b{\"u}chi definable tree languages.
\newblock In {\em Annual Conference for Computer Science Logic}, 2013.

\bibitem{Guidable}
Thomas Colcombet and Christof L{\"o}ding.
\newblock The non-deterministic mostowski hierarchy and distance-parity
  automata.
\newblock In {\em International Colloquium on Automata, Languages and
  Programming}, 2008.

\bibitem{GameAutomata}
Alessandro Facchini, Filip Murlak, and Michal Skrzypczak.
\newblock Rabin-mostowski index problem: A step beyond deterministic automata.
\newblock In {\em 2013 28th Annual ACM/IEEE Symposium on Logic in Computer
  Science}, pages 499--508, 2013.

\bibitem{RegisterGames}
Karoliina Lehtinen.
\newblock A modal mu perspective on solving parity games in quasi-polynomial
  time.
\newblock In {\em Proceedings of the 33rd Annual ACM/IEEE Symposium on Logic in
  Computer Science}, LICS '18, page 639–648, New York, NY, USA, 2018.
  Association for Computing Machinery.

\bibitem{lodingHDR}
Christof L{\"o}ding.
\newblock Logic and automata over infinite trees.
\newblock {\em Habilitation, RWTH Aachen, Germany}, 2009.

\bibitem{BorelDeterminacy}
Donald~A. Martin.
\newblock Borel determinacy.
\newblock {\em Annals of Mathematics}, 102(2):363--371, 1975.

\bibitem{NS21}
Damian Niwinski and Michal Skrzypczak.
\newblock On guidable index of tree automata.
\newblock In Filippo Bonchi and Simon~J. Puglisi, editors, {\em 46th
  International Symposium on Mathematical Foundations of Computer Science,
  {MFCS} 2021, August 23-27, 2021, Tallinn, Estonia}, volume 202 of {\em
  LIPIcs}, pages 81:1--81:14. Schloss Dagstuhl - Leibniz-Zentrum f{\"{u}}r
  Informatik, 2021.

\bibitem{NW05}
Damian Niwinski and Igor Walukiewicz.
\newblock Deciding nondeterministic hierarchy of deterministic tree automata.
\newblock In Ruy J. G.~B. de~Queiroz and Patrick C{\'{e}}gielski, editors, {\em
  Proceedings of the 11th Workshop on Logic, Language, Information and
  Computation, WoLLIC 2004, Fontainebleau, France, July 19-22, 2004}, volume
  123 of {\em Electronic Notes in Theoretical Computer Science}, pages
  195--208. Elsevier, 2004.

\bibitem{SW16}
Michal Skrzypczak and Igor Walukiewicz.
\newblock Deciding the topological complexity of b{\"{u}}chi languages.
\newblock In Ioannis Chatzigiannakis, Michael Mitzenmacher, Yuval Rabani, and
  Davide Sangiorgi, editors, {\em 43rd International Colloquium on Automata,
  Languages, and Programming, {ICALP} 2016, July 11-15, 2016, Rome, Italy},
  volume~55 of {\em LIPIcs}, pages 99:1--99:13. Schloss Dagstuhl -
  Leibniz-Zentrum f{\"{u}}r Informatik, 2016.

\end{thebibliography}

\end{document}